\documentclass[a4paper]{jpconf}
\usepackage{graphicx}
\begin{document}
\title{Neutrino Experiments: \\ Status, Recent Progress, and Prospects}

\author{Steve Brice}

\address{Fermilab, Neutrino Department, Mail Station 309, Batavia, IL 60510-500, USA}

\ead{sbrice@fnal.gov}

\begin{abstract}
Neutrino physics has seen an explosion of activity and new results in the last decade. In this report the current state of the field is summarized, with a particular focus on progress in the last two years. Prospects for the near term (roughly 5 years) are also described.
\end{abstract}

\section{Introduction}

Neutrinos have been a focus of experimental effort over the last decade with old questions answered and new ones emerging. This review attempts to summarize the state of the field, highlighting progress made in the last two years (the time since the last EPS HEP conference), and outlining prospects for the near future.

Of necessity many neutrino related areas have been omitted from this work, either because they are covered in other parts of the EPS HEP conference, because they are too futuristic to be considered prospects for the near term, or because time and page constraints do not allow complete coverage. In particular, theoretical developments and the physics of cosmic ray neutrinos are covered in other reviews. Neutrinos in cosmology, neutrino magnetic moment searches, and Big Bang relic neutrinos are absent due to lack of time and space. There is also no mention of Superbeam experiments beyond NO$\nu$A and T2K and no mention of neutrino factories or beta beams as these subjects will not be real in the near ($\sim$5 year) term.

\begin{figure}[b]
\begin{center}
\includegraphics[width=\textwidth]{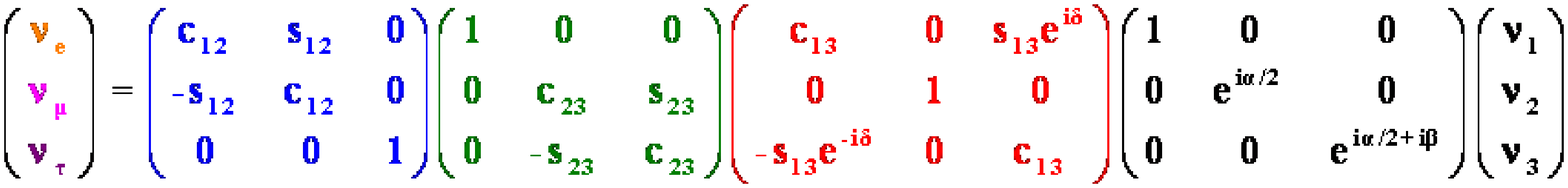}
\caption{The matrix transformation from neutrino mass eigenstates to neutrino flavor eigenstates. The shorthand ``$c_{13}$ stands for ``$\cos\theta_{13}$ and similarly for $s_{13}$ etc. $\alpha$ and $\beta$ are the two Majorana phases and $\delta$ is the CP violating phase.}
\label{fig:matrix}
\end{center}
\end{figure}

Over the last decade the theoretical framework with which we describe the three known neutrinos has crystallized. The three neutrino mass eigenstates, conventionally known as $\nu_1$, $\nu_2$, and $\nu_3$ are related to the three flavor eigenstates $\nu_e$, $\nu_\mu$, and $\nu_\tau$ by a unitary matrix that can be conveniently broken into four parts. These are shown in Fig.~\ref{fig:matrix}. This particular way of breaking up the transformation is logical on experimental grounds. Typically, a given experiment or measurement will be looking to study just one of these matrices or sectors, though knowledge of the other sectors is often needed to do this. One can then speak of measurements that probe the 12 sector, the 23 sector, the 13 sector, or the mass sector.

\begin{figure}
\begin{center}
\includegraphics[width=0.6\textwidth]{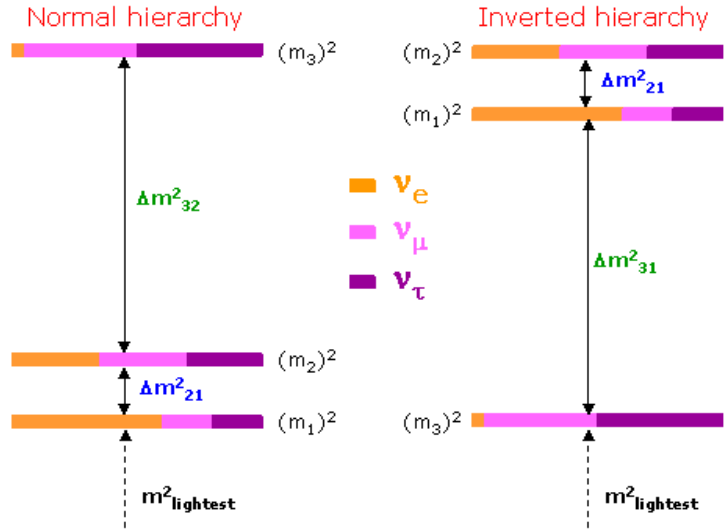}
\caption{The two possible mass orderings for the three known neutrino flavors.}
\label{fig:hierarchy}
\end{center}
\end{figure}

\begin{figure}[b]
\begin{center}
\includegraphics[width=0.4\textwidth]{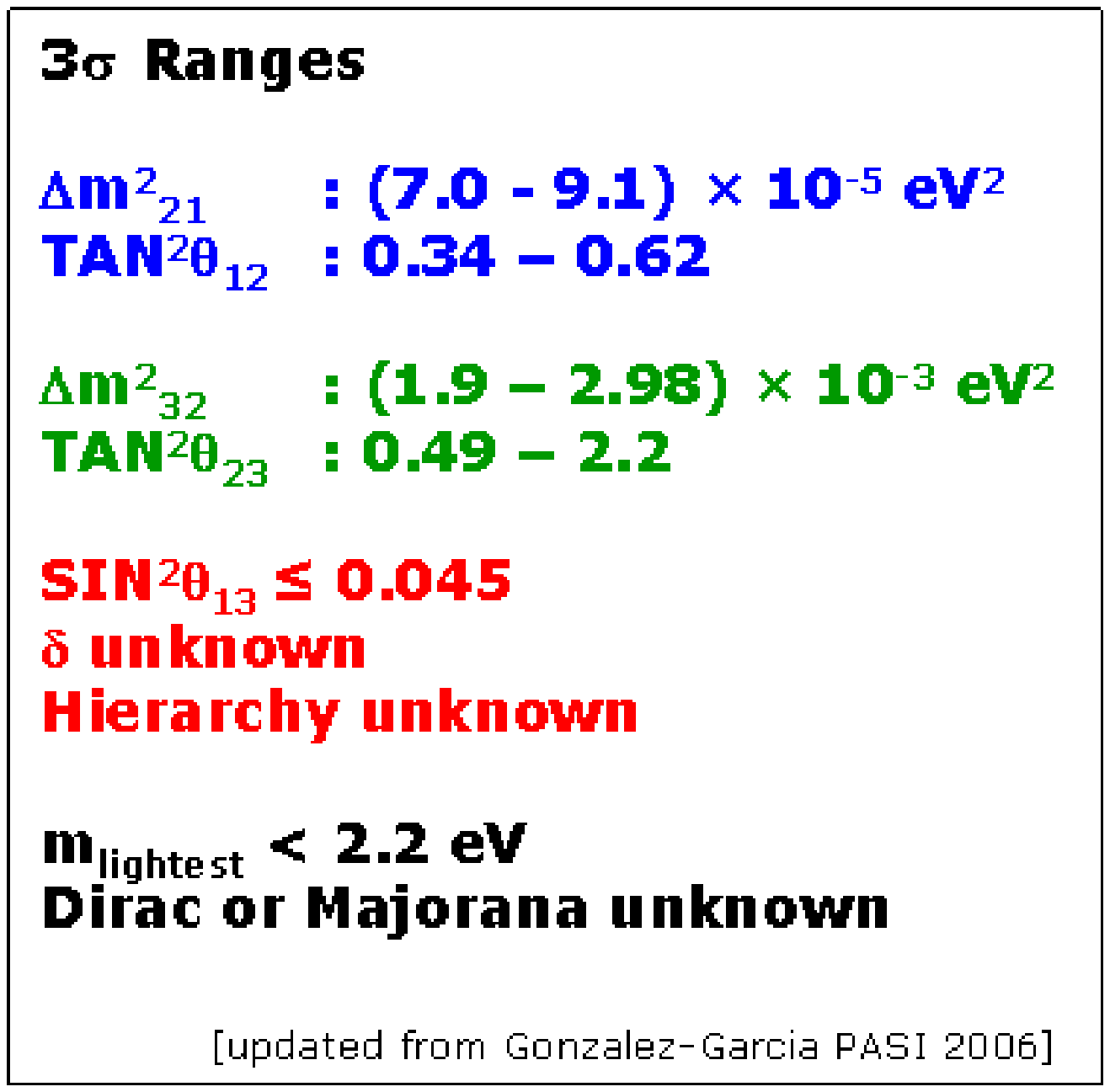}
\caption{The current state of knowledge of the neutrino parameters.}
\label{fig:numbers}
\end{center}
\end{figure}

Being an interference phenomenon, neutrino oscillation is sensitive to differences between squared neutrino masses rather than the masses themselves. Fig.~\ref{fig:hierarchy} displays graphically what oscillation experiments have so far taught us about the neutrino masses. The overall mass scale is not known, but the lightest neutrino is constrained by tritium beta decay measurements to be less than about 2.2 eV. The solar neutrino experiments and the KamLand reactor experiment measure the squared mass difference between the 1 and 2 mass eigenstates to be $(7.0-9.1) \times 10^{-5}$eV$^2$. The atmospheric neutrino measurements, as long baseline measurements of $\nu_\mu$ disappearance, constrain the squared mass difference between the 2 and 3 mass eigenstates to be $(1.9-2.98) \times 10^{-3}$eV$^2$. Both of these are 3$\sigma$ ranges. It is not determined, however, whether the mass of $\nu_3$ is larger or smaller than the $\nu_1$ and $\nu_2$ masses. These two scenarios are referred to as the normal and inverted hierarchy respectively. Fig.~\ref{fig:numbers} lists what we currently know about the neutrinos using the same color scheme established by Figs.~\ref{fig:matrix} and \ref{fig:hierarchy}. 

The structure of this review is provided by the structure of the transformation in Fig.~\ref{fig:matrix}. There are sections describing experiments that measure the 12 Sector, the 23 Sector, the 13 Sector, and, finally, the Mass Sector. Before entering this logical progression, however, there is a diversion in the next section to describe the new MiniBooNE results and how they probe the indication of oscillations from LSND.

\section{MiniBooNE and the LSND Result} \label{sec:miniboone}
The Liquid Scintillator Neutrino Detector (LSND) experiment \cite{lsnd} operated at Los Alamos National Lab in the 1990's and produced evidence for $\bar \nu_\mu \rightarrow \bar\nu_e$ oscillations at the $\Delta m^2 \sim 1$ eV$^2$ scale. Although the Karlsruhe Rutherford Medium Energy Neutrino Experiment (KARMEN) 
observed no evidence for neutrino oscillations \cite{karmen} in a similar mass range, a joint analysis \cite{joint} showed compatibility at 64\% CL. This $\Delta m^2$ scale is incompatible with those of the solar \cite{homestake98,sage99,gallex99,sno,sk02} and ``atmospheric'' \cite{kam,sk98,soudan99,macro01,k2k03,minos06} oscillations, and so requires there be more than 3 neutrinos if all three are to be interpreted as evidence of neutrino oscillation.

\begin{figure}
\begin{center}
\includegraphics[width=0.6\textwidth]{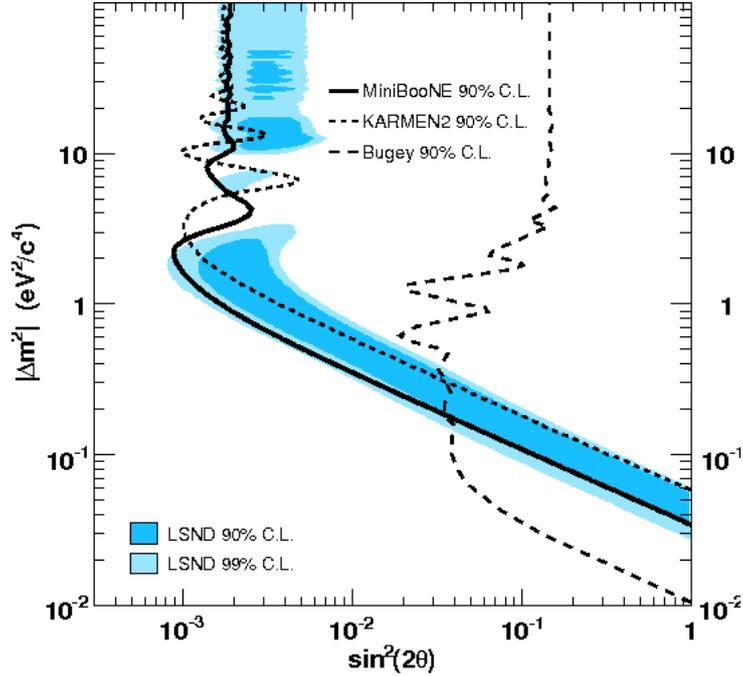}
\caption{The region of oscillation parameter space excluded at 90\% C.L. by the MiniBooNE result \cite{miniboone}. Also shown are the regions allowed by the LSND result \cite{lsnd} at 90\% C.L. (pale blue) and 95\% C.L. (dark blue), and the 90\% exclusion contours of the KARMEN2 \cite{karmen} and Bugey \cite{bugey} experiments.}
\label{fig:MBresult}
\end{center}
\end{figure}

The Mini-Booster Neutrino Experiment (MiniBooNE) was built to test the oscillation interpretation of the LSND result. It uses 8 GeV protons from the Fermilab Booster to produce a high purity beam of $\sim0.7$ GeV $\nu_\mu$ by running the protons into a Be target followed by a focusing horn. The detector is located 541m from the target and comprises a spherical tank of inner radius 610 cm filled with 800 tons of pure mineral oil (CH$_2$). The oil is viewed by 1280 8inch PMTs and surrounded by a veto region viewed by 280 PMTs. Using the pattern and timing of the Cerenkov and scintillation light hitting the tubes the experiment can distinguish electrons from other particles (in particular $\mu$s and $\pi^0$s) and so test for $\nu_\mu \rightarrow \nu_e$ oscillations. In April 2007 the experiment released it's first results \cite{miniboone}. The experiment found no evidence of neutrino oscillations in its analysis region above a neutrino energy of 475 MeV, though there was a excess of events found below this energy and this is currently under investigation. The exclusion plot that results from this measurement is shown in Fig.~\ref{fig:MBresult}. The MiniBooNE and LSND results are only compatible at the 2\% level if both are interpreted in the framework of two flavor neutrino oscillations. MiniBooNE is currently taking data in anti-neutrino mode (where the horn focusses negative particles) and intends to make a measurement of $\bar\nu_e$ appearance to more fully check the LSND result.

\section{The 12 Sector}
The 12 Sector comprises the mixing angle $\theta_{12}$ and the squared mass difference $\Delta m^2_{12}$ between $\nu_2$ and $\nu_1$. There are two types of experiment that have probed this sector; solar neutrino measurements and long baseline reactor anti-neutrino experiments. These two will be covered in the next two sections. 

\subsection{SNO and the Solar Neutrino Problem}

\begin{figure}
\begin{center}
\includegraphics[width=0.8\textwidth]{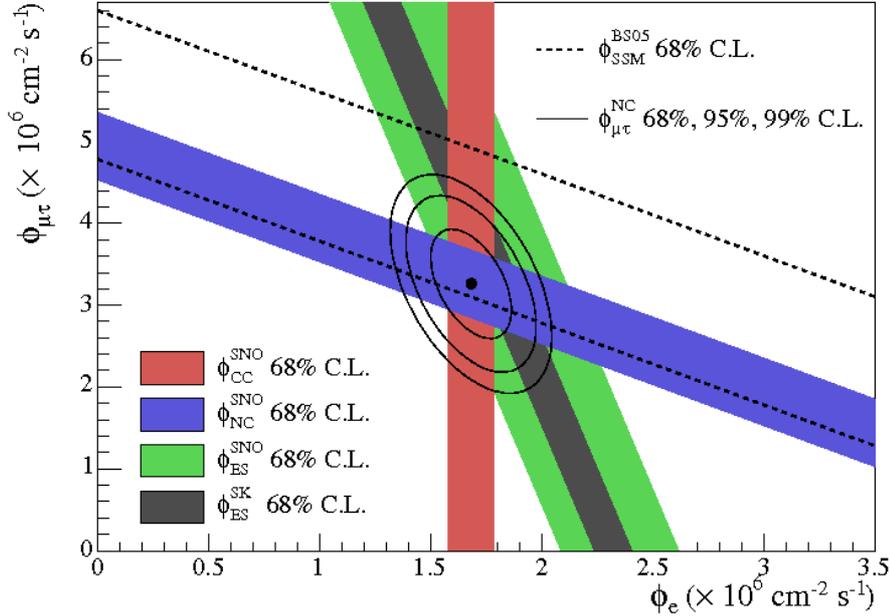}
\caption{The SNO results \cite{sno-salt} expressed as a measurement of the flux of electron neutrinos from the sun versus the flux of muon and tau neutrinos. The combination of SNOs Neutral Current and Charged Current channels (along with the SuperK measurement \cite{sk02} of neutrino electron scattering) show that the muon and tau flux is non-zero at the many sigma level.}
\label{fig:SNOresult}
\end{center}
\end{figure}

The values of $\Delta m^2_{12}$ and $\theta_{12}$ are such that the higher energy electron flavor neutrinos produced in the core of the sun leave it's surface as almost pure $\nu_2$ eigenstates. This is an effect of neutrino creation and adiabatic propagation through the very high matter densities in the sun. The expressions for the survival probability of an electron flavor neutrino are
\begin{equation}
   P( \nu_e \rightarrow \nu_e ) = \left\{ \begin{array}{cr} 
                            \sin^2\theta_{12} & E_\nu >\ \sim\!5\ {\rm MeV} \\
                     1 - \frac{1}{2}\sin^22\theta_{12} & E_\nu <\ \sim\!2\ {\rm MeV} 
                                         \end{array}\right.
\label{eqn:solar}
\end{equation}

Historically this probability for electron flavor neutrinos to disappear was known as the ``Solar Neutrino Problem'' and it's measurement started with the pioneering experiments of Ray Davis at the Homestake mine \cite{homestake98}. A whole sequence of measurements of solar neutrinos culminated in the Sudbury Neutrino Observatory (SNO) which used the deuterium in heavy water as a target for solar neutrinos. This enabled a measurement of both the electron neutrino flux using the electrons produced by the Charged Current interaction and the flux of all active neutrino flavors using the neutrons produced by the Neutral Current breakup of the deuteron. Fig.~\ref{fig:SNOresult} summarizes the beautiful set of measurements from SNO. It plots the measured flux of electron neutrinos against the measured flux of muon and tau neutrinos. The sun is not energetic enough to produce muons or taus and yet SNO shows conclusively that there is a flux of muon and tau neutrinos from the sun. The conclusion is that the electron flavor neutrinos are oscillating into muon and tau flavored on their way to the earth. The SNO measurements put the ``Solar Neutrino Problem'' to rest and, along with the KamLand reactor neutrino experiment, enable precision measurement of the 12 Sector parameters.

\subsection{KamLand} \label{sec:kamland}
If one solves the vacuum equation of motion using the matrices of Fig.~\ref{fig:matrix} and extracts the survival probability of electron neutrinos (or more strictly electron anti-neutrinos) one obtains
\begin{equation}
\begin{array}{l} 
 P( \bar\nu_e \rightarrow \bar\nu_e ) \approx 1 - \sin^22\theta_{12} \sin^2(1.27 \Delta m^2_{12} L/E_\nu ) \\ \\
 {\rm for}\ E_\nu/L = O(\Delta m^2_{12}),\ L{\rm (km)},\ E_\nu{\rm(MeV)},\ \Delta m^2_{12}{\rm(}10^{-3}{\rm eV}^2{\rm )}
\end{array}
\label{eqn:kamland}
\end{equation}
From Eqn.~\ref{eqn:kamland} it is clear that with the right choice for the neutrino energy $E_\nu$ and distance from neutrino source to detector $L$ an experiment could be sensitive to the parameters of the 12 Sector with an electron anti-neutrino disappearance measurement. 

\begin{figure}
\begin{center}
\includegraphics[width=0.6\textwidth]{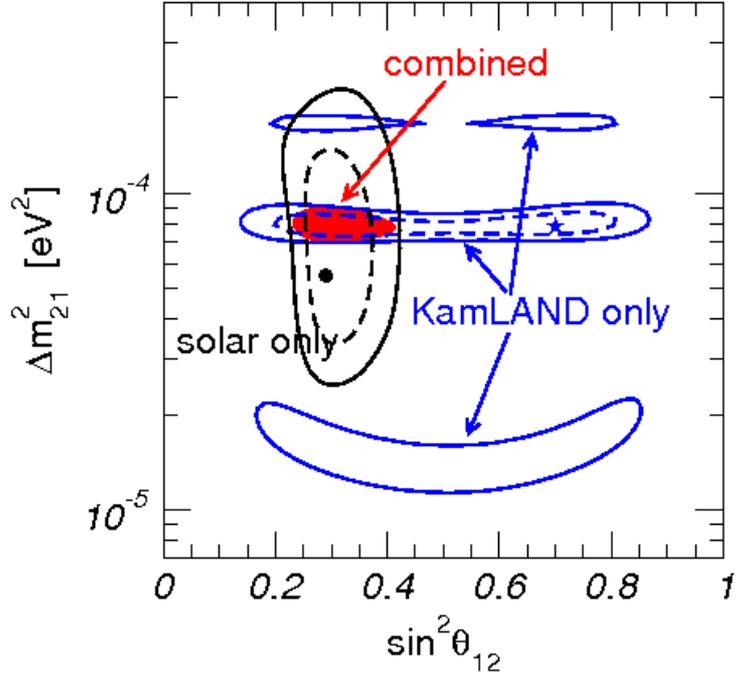}
\caption{Current state of knowledge of the 12 sector. Plot taken from \cite{schwetz}}
\label{fig:12sector}
\end{center}
\end{figure}

For reactor anti-neutrinos with energies of a couple of MeV the optimal distance from source to detector from Eqn.~\ref{eqn:kamland} is about 200km. This simple logic is what led to the proposal and construction of the KamLand experiment in Japan. The experiment uses 1000 tonnes of liquid scintillator in the Kamioka mine in central Japan to measure the flux of electron anti-neutrinos from all the power reactors in Japan. These reactors are at a flux weighted average distance of 180km, almost the ideal length. The KamLand experiment measured a deficit of electron anti-neutrinos consistent with that expected from the measurement of the 12 Sector parameters by SNO. In fact, as Fig.~\ref{fig:12sector} shows, the two experiments are wonderfully complementary. They use utterly different sources of neutrinos to measure the 12 Sector parameters via two completely different effects and the agreement is remarkable. 

\subsection{Upcoming 12 Sector Measurements}
The measurements of SNO and KamLand have put knowledge of the 12 Sector parameters on a very secure footing, nonetheless the parameters can always be better constrained and the energy dependent step in the survival probability evident in Eqn.~\ref{eqn:solar} has yet to be observed. The SNO experiment will soon publish the results of the third and final phase of measurement where discrete neutron counters (NCDs) were placed in the heavy water. This result should further pin down the mixing angle $\theta_{12}$. A new generation of experiments are attempting to measure solar neutrinos down to 1-2 MeV and look for the energy dependent shape of the survival probability: Borexino \cite{borexino} started data taking in mid 2007 and KamLand II, the Solar Neutrino phase of the KamLand experiment will soon get underway. Beyond these efforts there are several proposals for measurements of solar neutrinos below 1 MeV using the inverse $\beta$ decay of various isotopes.

\section{The 23 Sector}
The 23 Sector comprises the mixing angle $\theta_{23}$ and the squared mass difference $\Delta m^2_{32}$ between $\nu_3$ and $\nu_2$. Measurements of muon neutrino survival probability probe this sector. Using the matrices of Fig.~\ref{fig:matrix} and the vacuum equation of motion the probability that a muon neutrino of energy $E_\nu$ will survive as a muon neutrino after having traveled a distance $L$ is given by
\begin{equation}
\begin{array}{l} 
 P( \nu_\mu \rightarrow \nu_\mu ) \approx 1 - \sin^22\theta_{23} \sin^2(1.27 \Delta m^2_{32} L/E_\nu ) \\ \\
 {\rm for}\ L{\rm (km)},\ E_\nu{\rm(GeV)},\ m{\rm(eV)}
\end{array}
\label{eqn:numu_survival}
\end{equation}

In order to probe the mass ranges relevant to the 23 Sector with the GeV scale energies at which $\nu_\mu$s are most readily produced Eqn.~\ref{eqn:numu_survival} dictates that the baseline $L$ be in the few 100 km range. There are two types of experiments that can be done with these sorts of parameters: experiments measuring the neutrinos produced when cosmic rays hit the atmosphere and experiments located several hundred km from an accelerator source of muon neutrinos. The next two sections describe each of these types.

\subsection{SuperK Atmospheric Neutrinos and K2K}
Historically the disappearance of atmospheric muon neutrinos indicated by Eqn.~\ref{eqn:numu_survival} was known as the ``Atmospheric Neutrino Anomaly''. Several experiments \cite{soudan99,macro01}, but most notably the Kamiokande and SuperK experiments \cite{kam,sk98} in Japan, clearly showed that the flux of muon neutrinos from the atmosphere was significantly less than predicted whereas the flux of electron neutrinos agreed well with predictions (at least within the normalization systematic error). Furthermore, the zenith angle dependence, which is directly related to the baseline $L$, of the deficit is exactly what one would predict from Eqn.~\ref{eqn:numu_survival}. 

Since the atmosphere is not a very well controlled source of neutrinos a confirmation that the atmospheric results were due to the oscillations of Eqn.~\ref{eqn:numu_survival} was sought. The K2K experiment \cite{k2k03}, also in Japan, provided that reassurance. They used 12 GeV protons incident on a carbon target within a focusing horn to create a very pure beam of muon neutrinos at the KEK lab directed toward the SuperK detector $\sim\!250$ km away. They were able to demonstrate disappearance of the $\nu_\mu$s consistent with the predictions of Eqn.~\ref{eqn:numu_survival} with the parameters inferred from the atmospheric measurements. There was agreement both in the overall level of disappearance and in the spectral distortion produced by the energy dependence of the oscillation.

\subsection{MINOS}
\begin{figure}
\begin{center}
\includegraphics[width=0.8\textwidth]{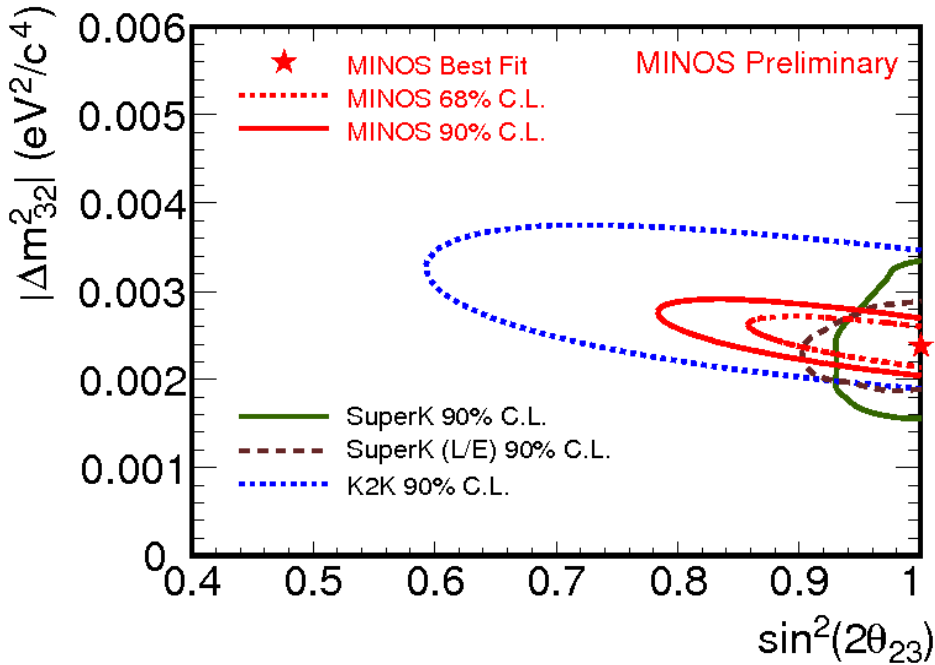}
\caption{Current state of knowledge of the 23 sector. Plot taken from \cite{minos}}
\label{fig:23sector}
\end{center}
\end{figure}
With the K2K experiment confirming the disappearance of muon neutrinos as a result of oscillations it falls to the MINOS experiment to make a precision measurement of the effect and further constrain the 23 Sector parameters. MINOS uses 120 GeV protons from the Fermilab Main Injector incident on a carbon target in a system of two horns to create a high purity $\nu_\mu$ beam that is directed toward the Soudan Mine 735 km away in northern Minnesota. There is a near detector at Fermilab and a far detector at Soudan. Each are made from alternating planes of steel and scintillator with the near detector weighing about 1 kton and the far detector 5.4 kton. The comparison of $\nu_\mu$ rates in near and far detectors and the very large numbers of protons incident on target enables a precision measurement of the $\nu_\mu$ disappearance to be made. New results from MINOS on the 23 Sector parameters were announced at the EPS HEP 2007 conference \cite{minos} and are shown in Fig.~\ref{fig:23sector} along with the older results from the SuperK atmospheric measurements and the K2K experiment. The oscillations in this sector are consistent with being maximal (i.e. $\theta_{23}=45^\circ$), but the precision with which $\theta_{23}$ is known is not great. It is an interesting open question how close to maximal $\theta_{23}$ really is.

\subsection{Near Term Future of the 23 Sector}
The recent MINOS results shown in Fig.~\ref{fig:23sector} are still statistics limited and, as MINOS continues to take data over the next few years, it will continue to improve the constraints on $\theta_{23}$ and $\Delta m^2_{32}$. The upcoming experiments T2K \cite{t2k} and NO$\nu$A \cite{nova} are designed to probe the 13 Sector and are described more fully in the later sections, but they will also be able to make very precise measurements of the 23 Sector and probe the issue of how close $\theta_{23}$ comes to being maximal. 

If the $\nu_\mu$s are oscillating away in experiments with GeV energies and baselines hundreds of kilometers long then they must be oscillating into tau neutrinos if the scheme of mixings and masses described in Figs.~\ref{fig:matrix},\ref{fig:hierarchy},\ref{fig:numbers} is correct. It would be a powerful test of the scheme to look for these $\nu_\tau$s and the OPERA experiment \cite{opera} is designed to do just that. OPERA is a hybrid emulsion and tracking detector that has recently started taking data in Gran Sasso. It detects neutrinos from the CNGS beam created at CERN. The baseline is 732 km and the mean neutrino energy 17 GeV. This energy is high enough for any $\nu_\tau$s in the beam to be able to create $\tau$ leptons via Charged Current interactions. OPERA will look for these $\tau$s by carefully scanning the 1.8 kton of lead/emulsion bricks that form the heart of the detector.

\section{The 13 Sector}

\begin{figure}
\begin{center}
\includegraphics[width=0.6\textwidth]{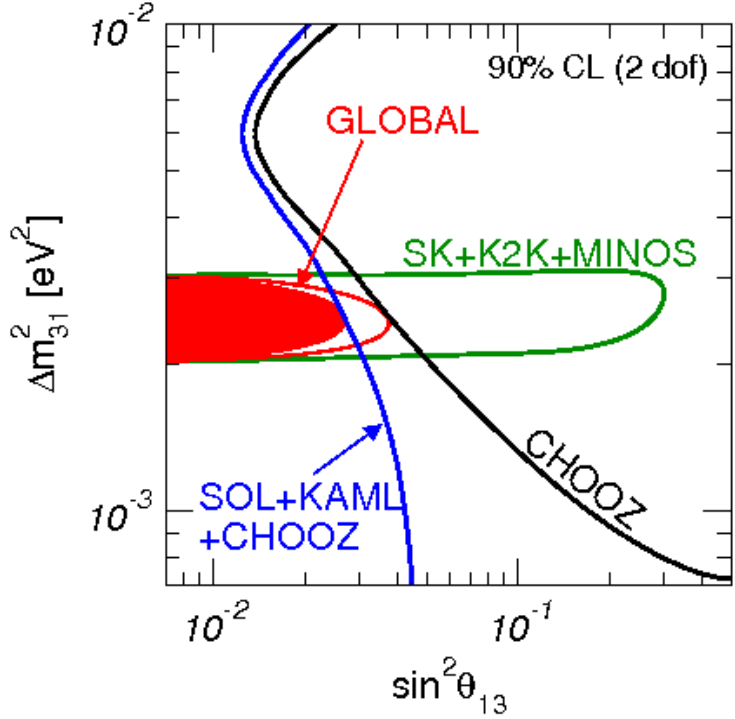}
\caption{Current state of knowledge of the 13 sector. Plot taken from \cite{schwetz}}
\label{fig:13sector}
\end{center}
\end{figure}

The 13 Sector comprises the mixing angle $\theta_{13}$, the phase $\delta$ which, if different from 0 or $\pi$, would induce CP violation into the scheme of neutrino oscillations, and the hierarchy which is just the question of whether the masses are ordered with the almost degenerate doublet $\nu_1$ and $\nu_2$ higher or lower than the $\nu_3$ mass. There are no constraints of any significance on $\delta$ and the hierarchy is unknown. We do have constraints on $\theta_{13}$, however, provided by experiments measuring electron anti-neutrino disappearance at baselines roughly 1km from the reactor source. The oscillation is described in the next section and the best of these measurements to date is the Chooz experiment that operated in France in the 1990's \cite{chooz}. Shown in Fig.~\ref{fig:13sector} are the current constraints on the 13 Sector mixing parameters. Note that the smallness of $\Delta m^2_{12}$ ensures that $\Delta m^2_{31}$ and $\Delta m^2_{32}$ are the same to the level of precision at which current and near future experiments operate.

Besides the disappearance of reactor anti-neutrinos the 13 Sector can also be probed by looking for $\nu_\mu \rightarrow \nu_e$ oscillations using baselines and energies sensitive to the parameters of both the 12 and 23 Sectors. Future experiments using this process are described in a later section.

\subsection{Reactor Neutrino Disappearance}
Sec.~\ref{sec:kamland} described how the KamLand experiment used reactor anti-neutrinos and a baseline of order 200 km to measure parameters of the 12 Sector. In this configuration the $L/E_\nu$ is matched to $\Delta m^2_{12}$. By instead measuring the disappearance of reactor anti-neutrinos at a baseline $L$ of $\sim1$km one can match the $L/E_\nu$ to $\Delta m^2_{31}$ and be sensitive to the parameters of the 13 Sector. The expression for the electron anti-neutrino survival probability is then
\begin{equation}
\begin{array}{l} 
 P( \bar\nu_e \rightarrow \bar\nu_e ) = 1 - \cos^4\theta_{13}\sin^22\theta_{12}\sin^2\Delta_{21}
 - \sin^22\theta_{13} \left( \cos^2\theta_{12}\sin^2\Delta_{31} + \sin^2\theta_{12}\sin^2\Delta_{32} \right) \\ \\
 \Delta_{ij} \equiv 1.27\Delta m^2_{ij} L/E_\nu\ \ \ L{\rm (km)},\ E_\nu{\rm(MeV)},\ m{\rm(}10^{-3}{\rm eV)}
\end{array}
\label{eqn:reactor}
\end{equation}

The Chooz experiment used a single detector located 1.05 km from the reactor core. A new generation of these kilometer baseline reactor experiments is being constructed in the hope of improving on the Chooz measurement and further constraining the value of $\theta_{13}$, either limiting it to be even closer to zero or measuring a non-zero value for it. These new experiments hope to be sensitive to a value of $\sin^22\theta_{13}$ as small as 0.01. To do this they are making several upgrades to the Chooz approach. Most importantly they are using multiple detectors to cancel systematics. These detectors will be larger, be located at very high flux reactors, and be exposed to the beam for longer. They will be underground to reduce the effect of cosmics and be thoroughly calibrated. The main contenders in this next round are Double Chooz \cite{double_chooz}, located at the same site as the original Chooz experiment, and the Daya Bay experiment \cite{daya_bay} located in China. Both experiments are under construction and have a staged approach to the measurement, bringing out results first with a subset of the final detector configuration and later with the full complement of detectors.

\subsection{Long Baseline Electron Neutrino Appearance}
The second way to probe the 13 Sector is by looking for $\nu_\mu \rightarrow \nu_e$ oscillations at values of $L/E_\nu$ matched to $\Delta m^2_{32}$. One is then looking at a process usually used to probe the 12 Sector, but with experiment parameters appropriate to the 23 Sector. It is via this interference that the 13 sector is probed. The full expression for the oscillation probability is
\begin{eqnarray} 
\lefteqn{P( \nu_\mu \rightarrow \nu_e ) = } \nonumber \\
 &   & \sin^2\theta_{23} \sin^22\theta_{13} \frac{\sin^2(\Delta_{31}\mp aL)}{(\Delta_{31}\mp aL)^2} \Delta^2_{31}
 \  + \  \cos^2\theta_{23} \sin^22\theta_{12} \frac{\sin^2(aL)}{(aL)^2} \Delta^2_{21} \nonumber \\
 & + & \cos\delta \sin2\theta_{23} \sin2\theta_{12} \sin2\theta_{13} \cos\Delta_{32} \left(\frac{\sin(\Delta_{31}\mp aL)}{(\Delta_{31}\mp aL)} \Delta_{31} \right) \left( \frac{\sin(aL)}{(aL)} \Delta_{21} \right) \nonumber \\
 & + & \sin\delta \sin2\theta_{23} \sin2\theta_{12} \sin2\theta_{13} \sin\Delta_{32} \left(\frac{\sin(\Delta_{31}\mp aL)}{(\Delta_{31}\mp aL)} \Delta_{31} \right) \left( \frac{\sin(aL)}{(aL)} \Delta_{21} \right) \nonumber \\
\label{eqn:nueapp}
\end{eqnarray}
\begin{displaymath}
\Delta_{ij} \equiv 1.27\Delta m^2_{ij} L/E_\nu,\ \ \ L{\rm (km)},\ E_\nu{\rm(GeV)},\ m{\rm(eV)},\ \ \ 
a \equiv G_F N_e / \surd2 \approx (4000{\rm\ km})^{-1}
\end{displaymath}

Eqn.~\ref{eqn:nueapp} is not a simple expression. With the exception of the overall neutrino mass scale, it involves all the parameters defined in Figs.~\ref{fig:matrix},\ref{fig:hierarchy},\ref{fig:numbers} in complicated combinations. The $\mp$ sign is negative if the hierarchy is normal and positive if it is inverted. When running in anti-neutrino mode and looking for $\bar\nu_\mu \rightarrow \bar\nu_e$ oscillations the same expression holds, but with the sign of the CP violating phase $\delta$ and the sign of the matter term $a$ flipped. The comparison of the oscillation in neutrino and anti-neutrino mode can therefore be used to search for CP violation, but the complicating effects of the matter potential have to be understood. Two of the three terms on the right hand side of Eqn.~\ref{eqn:nueapp} depend on $\sin2\theta_{13}$ and enable the search for $\nu_\mu \rightarrow \nu_e$ oscillations to look for a non-zero $\theta_{13}$.

There are two experiments under construction that will use Eqn.~\ref{eqn:nueapp} to probe the 13 Sector, first by searching for a non-zero $\theta_{13}$ and if one is found then determining the hierarchy and searching for CP violation. The two experiments are NO$\nu$A and T2K. NO$\nu$A will use the beamline currently used by MINOS, with a new detector being built 810km away in far northern Minnesota. T2K will use the existing SuperK detector and the beam will be sent from the JPARC accelerator lab currently under construction about 250 km away. Both experiments will make use of an off-axis beam to reduce backgrounds and increase sensitivity. As one moves off the beam axis of a conventionally produced neutrino beam the high energy tail of neutrinos is dramatically reduced leading to fewer backgrounds from Neutral Current events. Also, the $\nu_e$ contamination of the beam from sources such as $K^+$ and $\mu^+$ decay is significantly reduced. These two features come at little cost in $\nu_\mu$ flux, in fact the more peaked off-axis $\nu_\mu$ spectrum can be better tuned to the oscillation maximum.

\subsection{GeV Neutrino Cross-Sections}

\begin{figure}
\begin{center}
\includegraphics[width=0.6\textwidth]{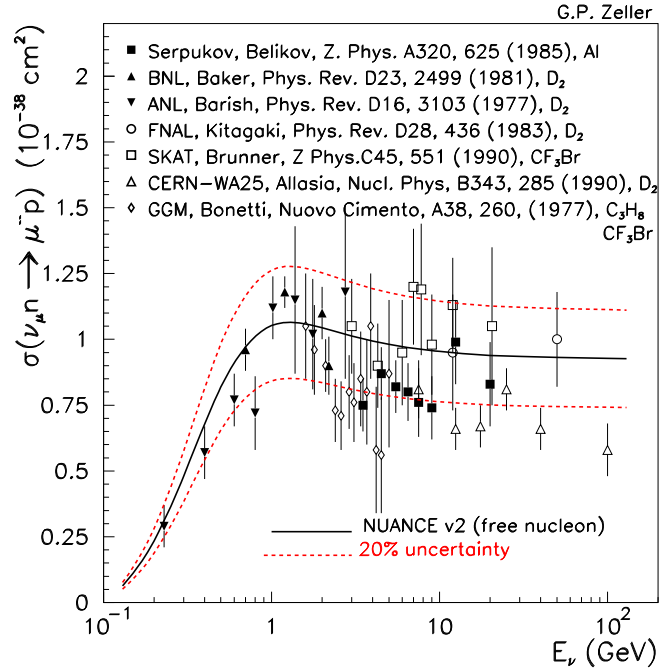}
\caption{A compilation of $\nu_\mu$ Charged Current Quasi-Elastic cross-section measurements. Plot taken from \cite{ccqe-xsec}}
\label{fig:ccqe-xsec}
\end{center}
\end{figure}

On several occasions this review has noted that current or upcoming experiments are making measurements with neutrinos of GeV energies. Typically these experiments use the Charged Current Quasi-Elastic (CCQE) interaction because the neutrino energy and flavor can be inferred from the final state particles. Fig.~\ref{fig:ccqe-xsec} is a compilation of the world's CCQE cross-section measurements in the few GeV range. It is clear that the CCQE cross-section in this range is not particularly well know. For experiments with a near detector this level of cross-section uncertainty is tolerable as it largely cancels when one takes a ratio of near detector to far detector rates (so long as the two detectors have the same target nuclei). The situation is more serious for the cross-sections of processes that can produce backgrounds to either a search for $\nu_e$ appearance or $\nu_\mu$ disappearance. Neutral Current $\pi^0$ production and Charged Current $\pi^+$ production are two examples where the cross-sections are very poorly known. Even using near to far ratios, if the upcoming experiments are to perform to their full capacity it will be necessary for the signal and background cross-sections to be better measured. To this end two current experiments, MiniBooNE and SciBooNE and one near future experiment, MINER$\nu$A are setting out to measure neutrino cross-sections in the GeV energy range.

MiniBooNE has been described in Sec.~\ref{sec:miniboone} in the context of its test of the LSND measurement. The experiment has recorded almost 800,000 neutrino interactions in both neutrino and anti-neutrino mode and is in the process of producing a whole range of cross-section measurements. SciBooNE is a new experiment that has taken the SciBar fine grained tracking calorimeter from the K2K experiment and put it into the same Booster Neutrino Beamline at Fermilab that MiniBooNE uses. The fine grained detector will give SciBooNE an ability to distinguish final states that big Cerenkov detectors like MiniBooNE cannot match. The detector is already taking data and first results are expected soon. The MINER$\nu$A experiment is building a high granularity detector in the NuMI beamline at Fermilab. It will have the ability to run with a variety of target materials. It has a large physics program including cross-section measurements in the few GeV energy range.

\section{Mass Measurements}
The mass sector comprises the mass of the lightest neutrino, the two Majorana phases in the mixing matrix ($\alpha$ and $\beta$ in Fig.~\ref{fig:matrix}) and the question of whether the neutrinos are Dirac or Majorana particles. Since the squared mass differences can be measured with the oscillation experiments already described it is only the overall mass scale, in the form of the lightest neutrino mass, that is left to be determined. This lightest mass can be constrained by kinematic measurements of the electron produced in $\beta$ decay. The masses, Majorana phases, and the Dirac or Majorana nature of the neutrino can be studied by looking for neutrinoless double $\beta$ decay. These two experiment types are described in the next two sections.

\subsection{Kinematic Mass Measurement}
The expression for the electron energy spectrum from an allowed $\beta$ decay is given in Eqn.~\ref{eqn:beta_decay}.
\begin{equation}
\frac{dN}{dE} = C F(Z,E) pE(E_0-E) \sum_i |U_{ei}|^2 \left[(E_0-E)^2-m_i^2\right]^\frac{1}{2} \Theta(E_0-E-m_i)
\label{eqn:beta_decay}
\end{equation}
where $C$ is a constant, $F(Z,E)$ is the Fermi function, $p$ and $E$ are the electron momentum and energy, $E_0$ is the electron endpoint energy, $U_{ei}$ is the element of the mixing matrix connecting the electron flavor neutrino to the $i$th neutrino mass eigenstate, and $m_i$ is the mass of the $i$th mass eigenstate. The presence of the neutrino masses $m_i$ in this expression only has a noticeable impact close to the endpoint energy $E_0$ where they cause the spectrum to cutoff at lower energy than would be expected if the neutrinos were massless. It is this features than enables experiments to search for neutrino mass by studying this endpoint region with extreme precision. 

The fraction of $\beta$ decays where the electron is close to the endpoint energy scales inversely with that endpoint energy and so the best $\beta$ decay limits on neutrino mass come from looking at the decays with the lowest endpoint energy. As well as having a very low endpoint energy of 18.6 keV, tritium is very light and so has nuclear and atomic corrections to the spectrum that are relatively easy to calculate. This is why the best limits on neutrino mass from $\beta$ decay kinematics come from measuring tritium. The current limit has the lightest neutrino mass at less than 2.2 eV with 95\% confidence \cite{tritium_limit}. The Katrin experiment \cite{katrin}, current under construction in Germany, is designed to have sensitivity to a lightest neutrino mass an order of magnitude smaller than this current limit. To do this the experiment will take the best aspects of the previous searches and combine them with improvements in several key areas. To improve statistical precision the tritium source will be $\sim\!80$ times stronger than previously used and the measuring period will be increased from about 100 days to something closer to 3 years. The spectrometer will have a better energy resolution ($\Delta E=0.93$ eV) and many of the systematics, like electron energy loss, that affected previous searches will be better controlled. It is hoped the experiment will start regular data taking in 2009.

\subsection{Neutrinoless Double $\beta$ Decay}
\begin{figure}
\begin{center}
\includegraphics[width=0.8\textwidth]{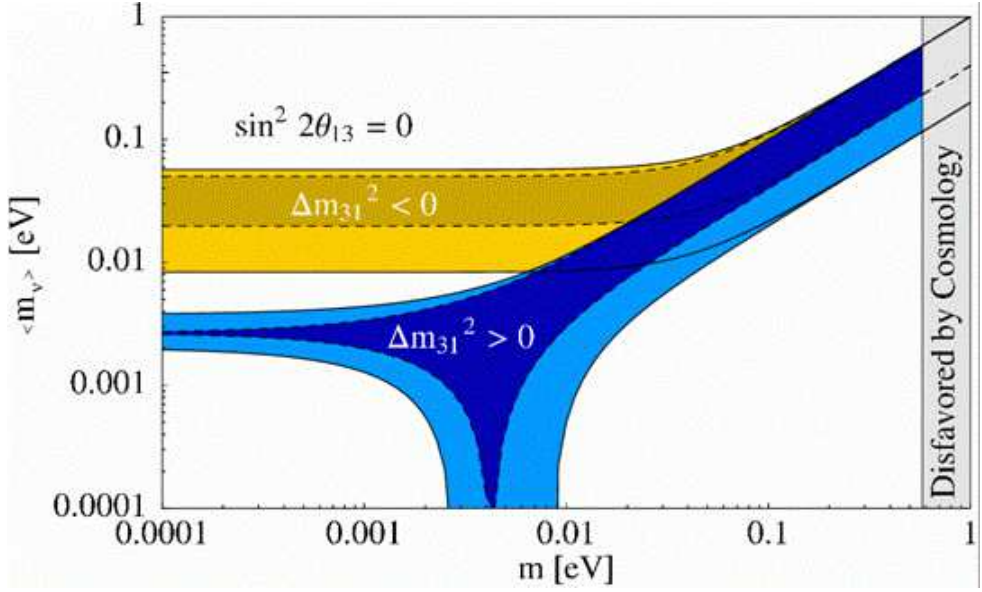}
\caption{The $0\nu\beta\beta$ effective mass as a function of the lightest neutrino mass. Plot taken from \cite{rodejohann}}
\label{fig:double_beta}
\end{center}
\end{figure}

In several nuclei with an even number of neutrons and an even number of protons the extra binding energy produced by the pairing leaves ordinary $\beta$ decay energetically forbidden. In such nuclei double beta decay, where two electrons are emitted, is left as the only viable decay mode. Two neutrino double beta (2$\nu\beta\beta$) decay has by now been observed in a number of nuclei, but neutrinoless double beta (0$\nu\beta\beta$) decay has yet to be convincingly seen. If observed, $0\nu\beta\beta$ decay would imply a massive Majorana neutrino. The rate for the process is given by
\begin{eqnarray}
\label{eqn:double_beta}
\Gamma_{0\nu} & = & G_{0\nu} |M_{0\nu}|^2 m_{\beta\beta}^2 \\
m_{\beta\beta} & = & \left| |U_{e1}|^2 m_1 + e^{i\beta}|U_{e2}|^2 m_2 +  e^{i\alpha}|U_{e3}|^2 m_3 \right| \nonumber 
\end{eqnarray}
where $G_{0\nu}$ is a readily calculable phase space factor and $M_{0\nu}$ is the, not so readily calculable, matrix element for the process. $\alpha$ and $\beta$ are the Majorana phases and the $U$s and $m$s are the mixing matrix elements and masses. The signature for the neutrinoless process is a peak in the measured energy of the pair of electrons at the Q value.

The search for $0\nu\beta\beta$ decay is quite mature with limits on the process pushed down to extremely tiny rates. Lifetimes for the ordinary beta decays of Uranium and Thorium isotopes are of the order the age of the universe. For the few observed $2\nu\beta\beta$ decays the measured lifetimes are about $10^{10}$ times the age of the universe, and the limits currently set on $0\nu\beta\beta$ decay are at about $10^{17}$ times the universe age. There is one recent claim for a detection of $0\nu\beta\beta$ decay in Germanium \cite{ge_claim}. The claim is controversial, but the experiment is the most sensitive to date, and if substantiated would imply an $m_\nu$ from Eqn.~\ref{eqn:double_beta} of 0.44 eV. The current generation of $0\nu\beta\beta$ experiments has this mass as a goal for their sensitivity.

One can combine information from the oscillation experiments and the $0\nu\beta\beta$ searches in an interesting way. Fig.~\ref{fig:double_beta} plots the $m_\nu$ measured by $0\nu\beta\beta$ decay and defined in Eqn.~\ref{eqn:double_beta} as a function of the lightest neutrino mass. If the mass hierarchy is normal then at least one neutrino must have a mass greater than $\surd(\Delta_{32}^2) \approx 0.05$ eV. If the hierarchy is inverted then at least {\em two} neutrinos must have masses greater than this value. These facts constrain the possible values that the $0\nu\beta\beta$ mass $m_\nu$ can have as a function of the lightest neutrino mass. In particular, if the hierarchy is inverted then $m_\nu$ must be greater than $\sim 0.01$ eV if the neutrino is a Majorana particle. The next generation of experiments have this mass scale as the goal for their sensitivity.

\section{Summary}
The last decade has been revolutionary in neutrino physics and the next decade promises an even more rapid development of our understanding. The masses and mixings being detected are giving us hints of physics well beyond the Standard Model. The question is whether we will be able to develop these hints into a path to a larger theory.


\section*{References}
\begin{thebibliography}{99}

\bibitem{lsnd}
C.~Athanassopoulos {\em et~al.}, 
Phys.\ Rev.\ Lett. 75, 2650 (1995);
77, 3082 (1996); 81, 1774 (1998);
A.~Aguilar {\em et~al.},
Phys.\ Rev.\ D 64, 112007 (2001).


\bibitem{karmen}
B.~Armbruster {\em et~al.},
Phys.\ Rev.\ D 65, 112001 (2002).

\bibitem{joint}
E.~Church {\em et~al.},
Phys.\ Rev.\ D 66, 013001 (2002).

\bibitem{bugey}
B.~Achkar {\em et~al.},
Nucl.\ Phys. B434, 503 (1995).

\bibitem{homestake98}
B.~T.~Cleveland {\em et~al.},
Astrophys.\ J. 496, 505 (1998).

\bibitem{sage99}
J.~N.~Abdurashitov {\em et~al.}, 
Phys.\ Rev.\ C 60, 055801 (1999).

\bibitem{gallex99}
W.~Hampel {\em et~al.},
Phys.\ Lett.\ B 447, 127 (1999).

\bibitem{sk02}
S.~Fukuda {\em et~al.},
Phys.\ Lett.\ B 539, 179 (2002).

\bibitem{sno}
Q.~R.~Ahmad {\em et~al.},
Phys.\ Rev.\ Lett. 87, 071301 (2001);
Q.~R.~Ahmad {\em et~al.},
Phys.\ Rev.\ Lett. 89, 011301 (2002);
S.~N.~Ahmed {\em et~al.},
Phys.\ Rev.\ Lett. 92, 181301 (2004).

\bibitem{kamland}
K.~Eguchi {\em et~al.},
Phys.\ Rev.\ Lett. 90, 021802 (2003);
T.~Araki {\em et~al.},
Phys.\ Rev.\ Lett. 94, 081801 (2005).

\bibitem{borexino}
The Borexino Collaboration,
C.~Arpesella {\em et~al},
arXiv:0708.2251v2 [astro-ph]

\bibitem{kam}
K.~S.~Hirata {\em et~al.}, 
Phys.\ Lett.\ B 280, 146 (1992);
Y.~Fukuda {\em et~al.},
Phys.\ Lett.\ B 335, 237 (1994).

\bibitem{sk98}
Y.~Fukuda {\em et~al.},
Phys.\ Rev.\ Lett. 81, 1562 (1998).

\bibitem{soudan99}
W.~W. M.~Allison {\em et~al.}, 
Phys.\ Lett.\ B 449, 137 (1999).

\bibitem{macro01}
M.~Ambrosio {\em et~al.},
Phys.\ Lett.\ B 517, 59 (2001).

\bibitem{k2k03}
M.~H.~Ahn {\em et~al.},
Phys.\ Rev.\ Lett. 90, 041801 (2003).

\bibitem{minos06}
D.~G.~Michael {\em et~al.},
Phys.\ Rev.\ Lett. 97, 191801 (2006).

\bibitem{miniboone}
A.~A.~Aguilar-Arevalo {\it et al.}  [The MiniBooNE Collaboration],
Phys.\ Rev.\ Lett.\  {\bf 98}, 231801 (2007)

\bibitem{sno-salt}
The SNO Collaboration 
Phys. Rev. C 72, 055502 (2005) 

\bibitem{schwetz}
T.~Schwetz
Phys.Scripta T127 (2006) 1-5 

\bibitem{minos}
The MINOS Collaboration
Contributed to 30th International Cosmic Ray Conference (ICRC 2007), Merida, Yucatan, Mexico, 3-11 Jul 2007,
arXiv:0708.1495 [hep-ex]

\bibitem{chooz}
CHOOZ collaboration, 
M. Apollonio et al., Phys. Lett. B 466 415.

\bibitem{double_chooz}
The Double Chooz Collaboration,
Proposal
arXiv:hep-ex/0606025

\bibitem{daya_bay}
The Daya Bay Collaboration
Proposal
arXiv:hep-ex/0701029

\bibitem{t2k}
The T2K Collaboration,
Letter of Intent
arXiv:hep-ex/0106019 

\bibitem{nova}
The NO$\nu$A Collaboration,
{\tt http://www-nova.fnal.gov/NOvA\_Proposal/Revised\_NOvA\_Proposal.html}

\bibitem{opera}
The OPERA Collaboration,
proposal available from
{\tt http://www.nu.to.infn.it/exp/all/opera/}

\bibitem{ccqe-xsec}
G. Zeller,
Contributed to the 2nd International Workshop on Neutrino-Nucleus Interactions in the Few GeV Region (NuInt02), Irvine, CA, December 2002,
arXiv:hep-ex/0312061v1

\bibitem{rodejohann}
M.~Lindner, A. Merle, W. Rodejohann
Phys.Rev. D73 (2006) 053005, 
hep-ph/0512143


\bibitem{tritium_limit}
C. Weinheimer et al., Phys. Lett. B460 (1999) 219; 
V.M. Lobashev et al., Proc. of the Int. Conf. Neutrino 2000, Sudbury, Canada, Nucl. Phys. B (Proc. Suppl.) 91 (2000) 280

\bibitem{katrin}
The KATRIN Collaboration
Letter of Intent (2001),
arXiv:hep-ex/0109033

\bibitem{ge_claim}
H.V. Klapdor-Kleingrothaus et al., Phys.Lett. B586, 198 (2004); 
H.V. Klapdor-Kleingrothaus et al., arXiv:hep-ph/0404062.


\end{thebibliography}
\end{document}